# Use of a Generic Identification Scheme Connecting Events and Detector Description in the ATLAS Experiment

C. Arnault, A. Schaffer
*LAL, 91898 Orsay FRANCE*

High energy physics detectors can be described hierarchically from the different subsystems to their divisions in r, phi, theta and to the individual readout channels. An identification schema that follows the logical decomposition of the ATLAS detector has been introduced allowing identification of individual readout channels as well as other parts of the detector, in particular detector elements. These identifiers provide a sort of "glue" allowing, for example, the connection of raw event data to their detector description for position calculation or alignment corrections, as well as fast access to subsets of the event data for event trigger selection. There are two important requirements on the software to support such an identification scheme. First is the possibility to formally specify these identifiers in terms of their structure and allowed values. And second is to generate different forms of the identifiers optimised in terms of access efficiency to information content, compactness or search key efficiency. We present here the generic toolkit developed in the context of the ATLAS experiment to primarily provide the identification of the readout channels and detector elements. The architecture of the toolkit is decomposed into three parts: an XML-based dictionary containing the formal specification of a particular range of identifiers, a set of various identifier classes (offering various level of compaction), and finally a set of "helper" classes, specific for each detector system, which serve as intermediaries between the dictionary and the identifier classes to create, manipulate and interpret the identifiers. This architecture will be described as well as the various applications of this identification scheme.

## 1. OVERVIEW

The data coming from a HEP detector requires access to its "detector description" information in order to be used. For example, one needs to calibrate the individual readout channel responses, calculate their positions and correct for misalignment. This can be done in many ways, but ultimately this relies on a key-lookup to match the readout data with its detector description data.

The solution that we have chosen in the ATLAS experiment consists of an identification scheme that follows the logical hierarchical structure of the detector. For example, our silicon strip detector is composed of a barrel and two end-caps, and several layers of two-sided modules distributed in $\eta$ (the cylindrical z co-ordinate) and $\phi$. So we have formulated an identifier specification of each silicon strip as:

> Inner Detector / SCT / barrel or endcap / layer / phi_module / eta_module / side / strip

The hierarchical structure allows one to extract identifiers with the full or partial hierarchy – readout ids correspond to the full hierarchy. In ATLAS we have found the following "three-level" model useful:

> detector subsystem → detector elements → readout channels where for the silicon detector one would have SCT → wafer (side) → strip. Thus we are interested in "projecting out" from the specification the identifiers for the detector subsystem, detector elements and readout channels. This three-level model is reflected in both the event and detector description models. Identifiers are used as look-ups both within and between these models.

The software infrastructure to support this identification system consists of:

> an identifier dictionary that allows one to capture and query different identifier specifications and to extract various forms of identifiers,
>
> various identifiers, e.g. expanded, compact or hash, and
>
> identifier 'helpers" that are specific to each detector system and simplify the interactions with the dictionary in terms of creation/manipulation of the identifiers

The basic infrastructure has been in place since 1998. A complete specification of the identifiers for each detector system was completed in 2001, using a first implementation with expanded identifiers[1]. Since 2002, we have been migrating to the use of an identifier dictionary, which allows the use of compact and hash identifiers that are more efficient in terms of space and look-up speed.

## 2. IDENTIFIER AND RANGE CLASSES

We are currently using three forms of identifiers:

> Expanded – internal representation is a vector<short int>
>
> Compact – 32 bit representation, the values of each level are bit packed
>
> Hash – 32 bit representation, transformation of a set of compact ids to numbers from 0 to N-1, allows constant-time table look-up

---

[1] In the form of std::vector<short>, i.e. a short for each level of an identifier.

The expanded form is primarily used internally within the infrastructure classes, whereas the latter two forms are widely used by clients in the ATLAS offline software. The compact id, which contains the information of the identifier in a compact form, can be used as a key in a binary look-up. Hash ids have been introduced to reduce this to a constant-time look-up. The hash ids are "optimal" in the sense that they take the values of 0 to N-1 for a specific set of N identifiers. This is possible when one works within a certain "context", and is facilitated by the identifier dictionary that allows one to enumerate all possible values for a particular context. For the example given above, the set of detector elements for a given sub-detector can be enumerated and each assigned its "hash" id that is simply a number from 0 to N-1 for N elements. One may then use this hash id in a table look-up.

We have introduced two classes to "capture" an identifier specification. A Range class contains the allowed values at each level for a region of valid identifiers. For each level, the Range maintains either the minimum and maximum or an enumeration of values, where min/max can be wild-carded. For example, a region of valid identifiers for the SCT example above, can be given as:

2 / 2 / -2, 2 / 0 : 8 / 0 : 51 / 0 /  0 : 1 / 0 : 767

which means : Inner Detector / SCT / both endcaps / layers 0 to 8 / phi_modules 0 to 51 / eta_module 0 / sides 0,1 / strips 0 to 767

There is a MultiRange class that extends this kind of expression by specifying an "or-ed" expression of several individual ranges. It should be noted that while a Range always describes a contiguous set of values for each specified field, the MultiRange permits the specification of non-contiguous subsets. In general, a MultiRange captures the specification of identifiers for a complete sub-detector. The MultiRange provides a validity check for any identifier, and as well iterators over all identifiers in the specification.

## 3.  IDENTIFIER DICTIONARY

The identifier dictionary formally defines a logical hierarchy and provides a name for each of the hierarchy levels. The identifier dictionary describes the set of non-overlapping regions of valid values for the hierarchy levels, where each region corresponds to a Range object in memory and the full dictionary to a MultiRange object. The dictionary also allows the symbolic description of identifier fields (similar to C++ enumerated types, e.g. **barrel** or **endcap**). Finally, the regions may be divided into separate dictionaries for convenience, where a manager groups them together. This used for sub-detectors where clients tend to be interested in only a sub-set of identifiers belonging to a particular sub-detector. These sub-sets are manipulated by the id helpers (see below) and maintained by separate groups or people.

A dictionary has its primary description in the form of an XML file, which is modified by the maintainers. A DOM XML parser[2] converts this dictionary into a IdDictDictionary object with an API providing the follow operations:

Queries on regions or symbolic fields
Generation of a MultiRange object for a given selection
Generate a packed (32 bit) representation from an expanded identifier
Unpack the compact representation into an expanded identifier.

One can query the dictionary for the names of its levels or fields, or their possible values. The MultiRange generation can be for the whole dictionary, or one may select sub-set of regions and/or the desired depth of the identifier hierarchy.

The packing/unpacking operators follow an algorithm that minimises the binary representation. The basic algorithm works its way down the hierarchy specification and "re-uses" bits for non-overlapping sub-ranges. This has been combined with an "over-ride" mechanism where one may specify that a subset of regions should have a common bit-mapping. This allows efficient mask-and-shift operations to be performed within the sub-region.

Finally, there is a "tag" attribute to the **region** XML element that allows the specification and selection of different versions of the valid ranges of identifier values.

A (limited) example of a dictionary is:

```
<IdDictionary name="InnerDetector" >

 <!--
    Start by defining some symbolic labels used
    for some field (other fields will be specified
    by numeric ranges)
 -->

 <field name="part">
  <label name="Pixel" value="1" />
  <label name="SCT" value="2" />
  <label name="TRT" value="3" />
 </field>

 <field name="barrel_endcap">
  <label name="negative_endcap" value="-2" />
```

---

[2] We have worked with both XercesC and Expat parsers, and currently use Expat.

```xml
    <label name="negative_barrel"  value="-1" />
    <label name="barrel"           value="0"  />
    <label name="positive_barrel"  value="+1" />
    <label name="positive_endcap"  value="+2" />
  </field>

  <subregion name="SCT_barrel">
    <range field="part" value="SCT" />
    <range field="barrel_endcap" value="barrel" />
  </subregion>

  <subregion name="SCT_endcap">
    <range field="part" value="SCT" />
    <range field="barrel_endcap"
      values="negative_endcap positive_endcap" />
  </subregion>

  <subregion name="SCT_eta_module">
    <range field="wafer" minvalue="0"
      maxvalue="1" />
    <range field="strip" minvalue="0"
      maxvalue="767" />
  </subregion>

  <subregion name=
    "SCT_phi_negative_barrel_module">
    <range field="eta_module" minvalue="-6"
      maxvalue="-1" />
    <reference subregion="SCT_eta_module" />
  </subregion>

  <region>
    <reference subregion="SCT_barrel" />
    <range field="layer" value="0" />
    <range field="phi_module" minvalue="0"
      maxvalue="31" />
    <reference subregion=
      "SCT_phi_negative_barrel_module" />
  </region>

  <region>
    <reference subregion="SCT_barrel" />
    <range field="layer" value="1" />
    <range field="phi_module" minvalue="0"
      maxvalue="39" />
    <reference subregion=
      "SCT_phi_negative_barrel_module" />
  </region>
</IdDictionary>
```

The hierarchy levels are defined either explicitly as **field** XML elements, or implicitly as one defines the **region** XML elements levels. In the latter case, one defines the allowed values for each level with a **range** XML element, and the field name is one of its attributes. We have also introduced **subregion** XML elements for shared partial specification of the hierarchy that may be referenced by different regions.

**THJT008**

## 4. ID HELPERS

Identifier helper classes have been developed for each detector system to simplify the interactions with the dictionary. They localize the place in the software where identifiers may be created or decoded. This guarantees a coherent definition, which may change with the version of the geometry description, and allows for an evolution of the packing alrgoirthm.

Each helper is tailored to the specific detector identifiers, for example for decoding the specific fields. The helpers are initialized from their corresponding dictionary. It is the helpers that transform a set of compact identifiers into their corresponding hash identifiers, by simply enumerating the set, and they in general cache the table of identifiers for fast conversion.

The helpers define one or more sets of hash identifiers, for example for detector elements and readout channels as described above, by selection on which regions and/or to a specific depth in the hierarchy. The helpers can provide iterators over the set of identifiers. Finally, the helpers can provide fast access to neighbouring identifiers since this knowledge can be obtained from the dictionary, where one can specify that some hierarchy levels "wrap-around" in $2\pi$, for example for the cylindrical $\varphi$ coordinate.

## 5. IMPACT ON THE ATLAS DATA MODEL

Identifiers in ATLAS are used to identify offline software the individual readout channels[3] and detector elements, which correspond to groupings of readout channels. This has led ATLAS to introduce a two-level container for event data:

container → collection → T:Digit

where the collection granularity corresponds to the detector elements, which is defined differently for each sub-detector. The access to individual collections is done via the hash identifiers, which as well provide the connection to detector description information.

This model is currently being used by High Level Trigger studies (see [1]) where a fast selection is performed after a reconstruction within a region-of-interest (ROI). For example,

---

[3] Typically there is a separate online identification scheme corresponding to the electronics. The online ids are mapped to these "offline" ids fairly early on in the software chain.

ROI from level 1 trigger → region in ΔηΔφ → hash ids → collections to be decoded from online read-out the glue which connects the two models. The full deployment of compact and hash identifiers is currently on going. And better understanding of the performance issues will come in the near future.

## 6. CONCLUSIONS

We have presented here an identification scheme comprised of a specification language (based on XML) and an associated C++ toolkit. This is being used by the ATLAS experiment for generating and manipulating various forms of identifiers which are used throughout the event and detector description models, as well as acting as

**THJT008**